\documentclass[12pt,preprint]{aastex}
\usepackage{psfig,epsfig,amsfonts,amsmath,amssymb,graphicx,morefloats,appendix,tensor}
\usepackage{color}
\usepackage{natbib}
\begin{document}

\slugcomment{Accepted by ApJ: March 16, 2016}

\title{
Constraints on Planetesimal Collision Models in Debris Disks}

\author{Meredith A. MacGregor\altaffilmark{1}, 
David J. Wilner\altaffilmark{1}, 
Claire Chandler\altaffilmark{2}, 
Luca Ricci\altaffilmark{1}, 
Sarah T. Maddison\altaffilmark{3}, 
Steven R. Cranmer\altaffilmark{4}, 
Sean M. Andrews\altaffilmark{1}, 
A. Meredith Hughes\altaffilmark{5}, 
Amy Steele\altaffilmark{6}}
\altaffiltext{1}{Harvard-Smithsonian Center for Astrophysics,
  60 Garden Street, Cambridge, MA 02138, USA}
\altaffiltext{2}{National Radio Astronomy Observatory, Socorro, NM 87801, USA}
\altaffiltext{3}{Centre for Astrophysics \& Supercomputing, Swinburne University, Hawthorn, VIC 3122, Australia}
\altaffiltext{4}{Department of Astrophysical and Planetary Sciences, Laboratory for Atmospheric and Space Physics, University of Colorado, Boulder, CO 80309, USA}
\altaffiltext{5}{Department of Astronomy, Van Vleck Observatory, 
  Wesleyan University, Middletown, CT 06459, USA}
\altaffiltext{6}{Department of Astronomy, University of Maryland, College Park, MD 20742, USA}

\begin{abstract}

Observations of debris disks offer a window into the physical and dynamical properties of planetesimals in extrasolar systems through the size distribution of dust grains. In particular, the millimeter spectral index of thermal dust emission encodes information on the grain size distribution.  We have made new VLA observations of a sample of seven nearby debris disks at 9~mm, with $3\arcsec$ resolution and $\sim5$~$\mu$Jy/beam rms.  We combine these with archival ATCA observations of eight additional debris disks observed at 7~mm, together with up-to-date observations of all disks at (sub)millimeter wavelengths from the literature to place tight constraints on the millimeter spectral indices and thus grain size distributions. The analysis gives a weighted mean for the slope of the power law grain size distribution, $n(a)\propto a^{-q}$, of $\langle q \rangle = 3.36\pm0.02$, with a possible trend of decreasing $q$ for later spectral type stars. We compare our results to a range of theoretical models of collisional cascades, from the standard self-similar, steady-state size distribution ($q=3.5$) to solutions that incorporate more realistic physics such as alternative velocity distributions and material strengths, the possibility of a cutoff at small dust sizes from radiation pressure, as well as results from detailed dynamical calculations of specific disks.  Such effects can lead to size distributions consistent with the data, and plausibly the observed scatter in spectral indices.  For the AU Mic system, the VLA observations show clear evidence of a highly variable stellar emission component; this stellar activity obviates the need to invoke the presence of an asteroid belt to explain the previously reported compact millimeter source in this system.

\end{abstract}


\section{Introduction}
\label{sec:intro}

Debris disks represent the end stage of protoplanetary disk evolution.  
As such, they provide essential information on the processes of planet 
formation and circumstellar disk dispersion 
\cite[see reviews by][]{bac93,wya08,mat14}.  
The small dust grains detected at optical through centimeter wavelengths are 
thought to be produced by the collisional erosion of larger bodies, analogous 
to comets or Kuiper Belt Objects, commonly referred to as planetesimals.  
These kilometer-sized remnants of planet formation are effectively invisible
around other stars, 
but emission from the dusty debris produced in ongoing collisions offers a 
unique window into their physical properties and dynamics.  In particular, 
the spectral index of dust emission at millimeter to radio wavelengths 
encodes information on the grain size distribution within disks that can be 
used to constrain collisional models of planetesimals \citep{ric12,ric15b}.

The dominant mechanism responsible for stirring the planetesimals within 
debris disks to incite collisions remains controversial.  Stirring could be 
triggered by the ongoing formation of Pluto-sized bodies within the disk 
\citep{ken02,ken08} or by the dynamical effects of fully formed planets 
\citep{mus09}.  In either case, the reference model for dust production is the 
steady-state collisional cascade first formulated by \cite{doh69}.  This model 
assumes that the relative velocities and tensile strengths of the colliding 
bodies are independent of size, and leads to a power law size distribution, 
$n(a) \propto a^{-q}$ with index $q = 3.5$.  However, the fragmentation process
in debris disks could be more complex.  Recent analytic and numerical studies 
relax some of the restrictive assumptions of the reference model, and
incorporate more realistic dynamics and material physics.  Including a 
size-dependent velocity distribution predicts a steeper distribution, 
$q \sim 4$ \citep{pan12,gas12}, while decreasing the tensile strength of the 
colliding bodies predicts a shallower distribution, $q \sim 3$ \citep{pan05}. 

We present observations of seven debris disks at 9~mm using the Karl G. Jansky Very Large Array (VLA) of 
the National Radio Astronomy Observatory. These long wavelength observations probe emission
from the largest accessible dust grains in the disks. When combined with 
(sub)millimeter data, these observations provide a long lever arm in wavelength
that mitigates the impact of absolute calibration uncertainties on spectral 
index determinations. In addition, the spectral slopes at these long 
wavelengths are relatively insensitive to the effects of temperature, given
typical debris belts at 10's of K.  We combine this sample with 
observations of eight additional debris disks with the Australia Telescope Compact Array 
(ATCA) at 7~mm.  By pairing these long wavelength measurements with previous 
observations at shorter (sub)millimeter wavelengths, we can determine the 
spectral index of the dust emission and thus the grain size power law index 
$q$ for the combined sample of fifteen debris disks.  We compare our results 
with predictions from 
existing collisional cascade models and explore the effects of material 
strengths, velocity distributions, and small-size cutoffs on the steady-state 
grain size distribution.

In Section~\ref{sec:vla_sample} we present the VLA sample of debris disks.  Sections~\ref{sec:vla_obs} and \ref{sec:vla_results} discuss the VLA observations, analysis, and results.  In Section~\ref{sec:calc_q}, we describe how we determine the slope of the grain size distribution, $q$, and we present the results for the combined VLA and ATCA sample of debris disks.  In Section~\ref{sec:disc}, we compare our results to predictions from collisional cascade models and discuss results from analytical modeling of a steady-state grain population.  In Section~\ref{sec:concl}, we summarize the main conclusions of this study. 

\section{VLA Sample}
\label{sec:vla_sample}

We selected a sample of seven debris disks to observe with the VLA at 9~mm to 
measure spectral indices and constrain the slope of the grain size 
distribution, $q$.  The sample was assembled based on the following criteria: 
1)~accessible source declinations, $\delta > -35\degr$, 
2)~evidence in the literature for strong millimeter/submillimeter emission 
($F_\text{0.85mm} \gtrsim 8$ mJy), and 
3)~small enough angular extent to obtain reliable total flux measurements 
using the most compact array configurations.  
Table~\ref{tab:sample} lists the source positions and stellar properties, and a brief discussion of each target 
follows.  All of these disks have been studied extensively at other 
wavelengths and have well sampled spectral energy distributions (SEDs) through 
the far-infrared.  In addition, most of these disks have interferometric data 
that resolve their millimeter emission structure, either from the Submillimeter
Array (SMA) or the Atacama Large Millimeter/submillimeter Array (ALMA).  

\subsection{HD 377}
\label{subsec:hd377}

HD 377 is a G2V star at a distance of $39.1\pm2$ pc \citep{vanL07} with an 
estimated age of $\sim150$ Myr \citep{geer12}.  Spectral energy distribution 
modeling indicates the presence of dust between 3 and 150 AU \citep{roc09} 
and results in a two temperature component fit with a warmer inner belt at 
$T_\text{dust}\approx130$ K and a colder outer belt at 
$T_\text{dust}\approx50$ K \citep{mor11,panic13}.  \cite{cho15} recently detected HD 377 in reprocessed archival Hubble Space Telescope (HST) scattered light images. The disk was resolved at 870 $\mu$m with the SMA, 
revealing a symmetric belt of emission centered at $\sim47$ AU with a width of 
$\sim32$ AU \citep{stee15}.  \cite{gre12} did not detect the disk at $7.5-11.5$~mm with
the Green Bank Telescope (GBT), but the noise level of 14 $\mu$Jy/beam gave a 
$2\sigma$ upper limit on the disk flux of $<28$ $\mu$Jy.
No gas has been detected in the system \citep{geer12}.  
Searches with VLT/NACO have not revealed the presence of any gas giant planets with
masses between 3 and 7 $M_\text{Jup}$  and separations of $20-50$ AU \citep{apai08}. 

\subsection{49 Ceti}
\label{subsec:49ceti}

49 Ceti is an A1V star at a distance of $59\pm1$ pc \citep{vanL07} and a 
member of the Argus Association, indicating an age of $\sim40$ Myr 
\citep{tor08}.  The dust disk was resolved at 70 $\mu$m with 
\emph{Herschel/PACS} \citep{rob13}.  SED modeling indicates that this disk 
has two distinct components, a cold ($T_\text{dust}=62\pm1$~K) outer disk 
extending from 40 to 200 AU and a warmer ($T_\text{dust}=175\pm3$ K) inner 
belt within 40 AU \citep{wah07,hug08,rob13}.  New ALMA observations at 850 $\mu$m
are consistent with this picture and are best-fit by an inner belt of small dust grains
between $\sim4-60$~AU and an outer belt of larger grains between $\sim60-300$~AU (Hughes et al. in prep).

In addition to the dust disk, 49 Ceti is notable for exhibiting substantial
CO emission \citep{zuc95,den05,hug08}.  Resolved (SMA) observations of 
CO emission in the 49 Ceti system reveal that the inner disk is devoid of gas, 
while the outer belt contains $0.02\pm0.01$ $M_\oplus$ of gas \citep{hug08}.  
\emph{Herschel} spectroscopy indicates that this gas cannot be primordial, 
and may instead be secondary material coming from the destruction of comet-like 
ices \citep{rob13}.

\subsection{HD 15115}
\label{subsec:hd15115}

HD 15115 (``the blue needle'') is an F2V star at $45\pm1$ pc \citep{vanL07} 
whose space motions suggest membership in the $21\pm4$ Myr-old \citep{bin14} 
$\beta$ Pictoris moving group \citep{moor11}.  An infrared excess suggesting 
orbiting dust was noted in IRAS observations \citep{sil00}.  Subsequent 
scattered light imaging from the HST and other
telescopes have resolved an asymmetric, edge-on circumstellar disk 
\citep{kal07,deb08,rod12,maz14,sch14}.  Observations of 850 $\mu$m emission 
using the James Clerk Maxwell Telescope/SCUBA-2 suggested the presence of a 
reservoir of large dust grains in the disk, with low temperature 
($T_\text{dust} = 56\pm9$ K).  Observations at 1.3~mm with the SMA
resolve a belt of emission at $\sim110$ AU with a width of $\sim43$ AU 
\citep{mac15a}.  In addition, the millimeter emission shows a $\sim3\sigma$ 
feature aligned with the asymmetric western extension of the scattered light 
disk.  If real, this additional feature indicates that the distribution of 
larger grains in the disk may be asymmetric as well.

\subsection{HD 61005}
\label{subsec:hd61005}

HD 61005 (``the moth'') is a G8V star at a distance of $35\pm1$ pc 
\citep{vanL07}.  An argument has been made for membership in the Argus 
Association, suggesting an age of $\sim40$ Myr \citep{des11}.  The presence 
of dust was originally inferred from a significant Spitzer infrared excess 
\citep{carp05}, and follow-up HST images revealed a remarkable swept-back disk 
in scattered light that extends from $\lesssim10$ to 240 AU 
\citep{hines07,man09}.  The SED is best fit by a belt with 
$T_\text{dust} \approx 80$ K at $\sim90$~AU.
The disk was resolved at 1.3 mm with the SMA, and the continuum emission is 
fit by a narrow belt ($\Delta R/R = 0.05$) at $\sim70$ AU \citep{rica13,stee15}.
There is no indication that the dramatic scattered light asymmetry persists at 
millimeter wavelengths.  Given the wavelength-dependent nature of the belt 
morphology, it is possible that the observed swept-back features result from 
interactions with the interstellar medium, which would be expected to affect 
only smaller grains \citep{man09}.

\subsection{HD 104860}
\label{subsec:hd104860}

HD 104860 is an F8 zero age main sequence star ($\sim140$ Myr) at a distance of 
$48\pm2$ pc \citep{vanL07}.  The SED is well fit by a single temperature 
component at a radius of 105 AU and $T_\text{dust} = 33\pm3$ K 
\citep{naj05,roc09,paw14}.  The disk was resolved at 70, 100, and 160 $\mu$m 
with \emph{Herschel} \citep{mor13}, and these observations are best fit by 
including a second warm, less massive dust belt at $\sim5$ AU with 
$T_\text{dust} \approx 190$ K. SMA observations at 1.3 mm reveal an 
axisymmetric, broad belt at $\sim110$ AU with a width of $\sim100$ AU \citep{stee15}.
GBT observations at $7.5-11.5$~mm \citep{gre12} placed a $2\sigma$ upper limit on the
disk flux of $<24$ $\mu$Jy.  No gas has been detected in the system \citep{naj05}. 

\subsection{HD 141569}
\label{subsec:hd141569}

The HD 141569 system consists of HD 141569A, a young $\sim5$ Myr-old B9.5V star 
at $116\pm8$~pc \citep{vanL07}, and a pair of low mass comoving companions 
with spectral types M2 and M4 located $\sim7\farcs5$ away \citep{wei00}.  
Despite its young age, the star is surrounded by a highly evolved late-stage 
transition or early debris disk.  The morphology of this disk is complex. 
Scattered light and near-infrared imaging reveal asymmetric spiral structures 
between $\sim 175-210$~AU and $\sim300-400$~AU \citep{wei00,mou01,cla03,bil15}.
While this outer spiral structure is truncated at 175~AU, there is an 
additional inner debris belt between 10 and 50~AU (White et al. in prep.) 
with a dust temperature of $T_\text{dust}\approx 80$ K determined from SED 
modeling \citep{nil10}.

In addition to dust, this system contains between 20 and 460 $M_\oplus$ of gas 
\citep{zuc95,thi14}.  Most of this gas is associated with the outer part of 
the disk and distributed non-uniformly in two ring-like structures at 
$\sim90$ and 250~AU \citep{den05}.  However, near-infrared observations 
indicate that there is additional CO gas distributed between 10 and 50~AU, 
commensurate with the inner debris system \citep{bri02,goto06}.
Recent SMA and CARMA observations at 870~$\mu$m and 2.8~mm, respectively, resolve the gas disk (Flaherty et al., submitted)
and reveal a large inner hole in the CO gas distribution interior to $\sim29$~AU and an outer edge at $\sim224$~AU,
interior to the previously imaged scattered light rings.  Additionally, these observations
yield $T_\text{gas}\sim27$~K, lower than the dust temperature.

\subsection{AU Mic}
\label{subsec:aumic}

AU Mic is a nearby \cite[$9.91\pm 0.10$ pc;][]{vanL07} M1V star in the 
$\beta$ Pictoris moving group, suggesting an age of $21\pm4$ Myr \citep{bin14}.
The star is surrounded by a nearly edge-on circumstellar disk extending to a 
radius of at least 210 AU, discovered in coronographic images of scattered 
starlight \citep{kal04}.  ALMA Cycle 0 observations at 1.3 mm revealed (1) an 
outer belt with an emission profile that rises with radius out to 40 AU and 
(2) a newly recognized central peak that remained unresolved \citep{mac13}.  
The outer dust belt shows no evidence of asymmetries and is characterized by a 
dust temperature of $T_\text{dust} \approx 25$ K.  The central peak is 
$\sim6$ times brighter than the expected stellar photosphere, indicating an 
additional emission process in the inner region of the system, either a warm 
($T_\text{dust} \approx 75$ K) planetesimal belt or a hot stellar chromosphere 
or corona \citep{cra13,schu15}.  AU Mic is also well known to be a rich source 
of stellar flaring activity, notably at X-ray \citep{mit05,sch10} and 
ultraviolet \citep{rob01} wavelengths.

\section{VLA Observations}
\label{sec:vla_obs}

VLA observations for six of the seven sources in the sample were carried out between 
June and August 2014 at a wavelength of 9 mm (Ka band).  Two 120-minute scheduling blocks (SBs)
were observed in the D configuration (baseline lengths 0.04 to 1.03~km) for 
all disks except HD 141569, where only one SB was executed. HD 61005 was 
observed in the DnC configuration (with north-south baselines to 2.11~km, 
for its southern declination). We observed AU Mic, the seventh source, in June and July 2013 with
two 105-minute SBs in the D configuration and  one 105-minute SB in the 
C configuration (baseline lengths 0.05 to 3.38~km). 
Table~\ref{tab:obs} summarizes the essentials of these observations, including the observation dates, array configurations, number of antennas, baseline lengths, weather conditions, on-source time, and the gain calibrators 
used.  Overall, the weather conditions were very good for these summer 
observations (rms of $<6\degr$ measured with the Atmospheric Phase Interferometer at 11.7 GHz). The total bandwidth available for all observations was 8 GHz, 
split into $4 \times 2$ GHz basebands centered at 30, 32, 34, and 36 GHz.
The characteristic rms for these observations was $\sim5$ $\mu$Jy/beam and the typical 
natural weight beam FWHM was $\sim3\arcsec$.

The data from each track were calibrated separately using the CASA software 
package.  The passband shape was calibrated using available bright sources, 
mainly J0319+4130, J0609-1542, J1256-0547, and J1924-2914.  Observations of 
3C48 and 3C286 during each track were used to derive the absolute flux scale, 
with an estimated accuracy of $<10\%$.  Imaging and deconvolution were 
performed with the \texttt{clean} task in CASA (version 4.3.1).

\section{Results of the VLA Observations}
\label{sec:vla_results}

\begin{figure}[ht]
\centerline{\psfig{file=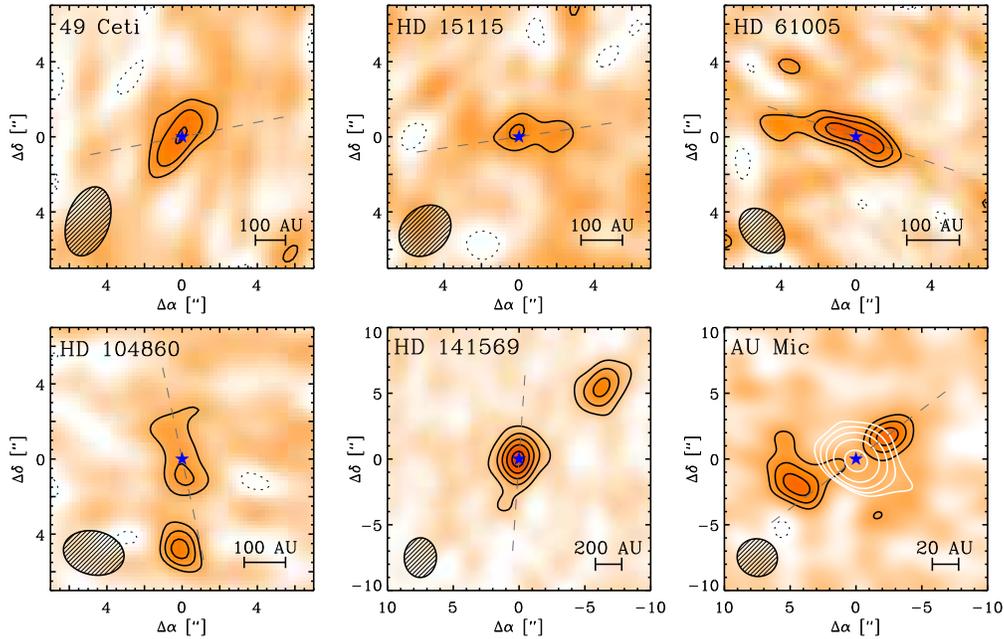,width=14cm,angle=0}}
\caption[]{\small Images of the 9 mm continuum emission from the six debris disks detected in the VLA sample.  Contour levels are in steps of $[2,3,4,6]$ $\times$ rms for all images, except HD 141569 (lower center) where contours are in steps of 3 $\times$ rms (characteristic rms $\sim5$ $\mu$Jy).  The white contours in the AU Mic image (lower right) mark the subtracted stellar component in steps of $[10,20,40,80,160]$~$\times$~rms.  The ellipse in each lower left corner indicates the synthesized beam size.  The star symbol marks the position of the stellar photosphere.  The dashed gray line indicates the position angle of the disk determined from previous optical, near-infrared, or millimeter imaging and listed in Table~\ref{tab:sample}.
}
\label{fig:fig1}
\end{figure}

Figure~\ref{fig:fig1} shows the VLA 9 mm images for the six detected debris disks in 
the sample.  HD 377 was undetected.  For the detected disks, the peak
signal-to-noise ratio achieved ranges between $\sim 4\sigma$ (HD 15115 and 
HD 104860) and $\sim 16\sigma$ (HD 141569).

The 49 Ceti and HD 141569 disks are unresolved by these observations.  
The HD 15115, HD 61005, and HD 104860 disks are all marginally resolved,
appearing extended in the direction of their position angles as determined 
from previous optical, near-infrared, and millimeter imaging.  Indeed, the HD 15115
disk shows a hint (at a $\sim2\sigma$ level) of the same asymmetric western 
extension seen in previous optical and millimeter imaging \citep{mac15a}.
For all five of these disks, the total flux density was determined by 
integrating the surface brightness over the area showing continuum emission 
at $\gtrsim2\sigma$ above the background rms in the image.  The total uncertainty
on the 9~mm flux density was taken to be the rms noise measured for the image added in 
quadrature with the $10\%$ uncertainty in the absolute flux calibration.
For HD 377, we obtain a $3\sigma$ upper limit on the total disk flux density.  
Table~\ref{tab:results} lists the synthesized beam size and position angle, the measured 
9 mm flux density, and the rms for all of the sources in the sample.  

AU Mic is the only disk that is well resolved in the sample.  
The C configuration affords high angular resolution (beam size $<1\arcsec$,
corresponding to $<$~$10$~AU at the distance of AU~Mic) relative to the
extent of the disk. This high resolution allows us to separate emission 
from the the central star and the disk. The star is very active at radio 
wavelengths, flaring on timescales shorter than the scheduling blocks.  
We were able to isolate the disk emission by subtracting a time-dependent 
point source model from the data to account for the stellar emission.  
In doing this, we could not avoid subtracting out some disk emission 
co-located with the star.  We estimated the total flux density as for the 
other disks in the sample.  However, the result obtained should be considered 
a lower limit to the disk emission.  A detailed account of the stellar 
emission observed in the AU Mic system is included in the appendix.

\section{Determining the Slope of the Grain Size Distribution $q$}
\label{sec:calc_q}

We adopt the method of \cite{ric12,ric15b} who used ATCA observations at 7~mm to constrain the millimeter grain size distribution of six debris disks.  Since the thermal dust emission from debris disks is optically thin, 
the flux density is given by $F_\nu \propto B_\nu(T_\text{dust})\kappa_\nu M_\text{dust}/d^2$, where $B_\nu(T_\text{dust})$ is the Planck function at the dust temperature $T_\text{dust}$, $\kappa_\nu\propto\nu^\beta$ is the dust opacity, expressed as a power law at long wavelengths, $M_\text{dust}$ is the total dust mass in the disk, and $d$ is the distance.  \cite{dra06} derived a relation between $\beta$, the dust opacity power law index, and $q$, the grain size distribution parameter: $\beta = (q-3)\beta_s$, where, $\beta_s$ is the dust opacity spectral index of small (i.e. much smaller than the observing wavelength) particles.  For size distributions that follow a power law with $3 < q < 4$ from blow-out grain sizes (on the order of $\sim\mu$m) to larger planetesimals ($\sim1 - 100$~km), $\beta_s = 1.8\pm0.2$, consistent with observations of both diffuse and dense interstellar clouds.
It should be noted, that given the restrictive assumption of $3<q<4$, this relation could prove inaccurate for any disks that have a size distribution index outside of this range.  Additionally, for dust compositions with large fractions of amorphous carbons or ices, values of $\beta_s$ can drop to $\sim1.4$ \cite[e.g. `cel800' produced by the pyrolysis of cellulose at $800\degr$~C in][]{dra06}, resulting in higher $q$ values for a given $\beta$.  For a representative case, where $\beta=0.5$, $q=3.28$ and $3.36$ for $\beta_s=1.8$ and $1.4$, respectively.  The change in $q$ due to a change in $\beta_s$ is $\sim0.08$ in this case, comparable to the uncertainties on $q$ derived by our analysis. Since we do not expect this effect to be large, and given our limited knowledge of the grain compositions in the debris disks in our sample, we assume $\beta_s = 1.8$ for the purposes of our analysis.    
This relation between $\beta$ and $q$ has been found to be very accurate for different dust models considered in the literature to interpret the millimeter wavelength emission of young circumstellar disks \cite[see e.g.][]{dal01,ric10a,ric10b}.

For debris disks around solar-type and earlier stars, the dust is typically 
warm enough ($k_BT_\text{dust} >>h\nu$) for the Planck function at long 
wavelengths to reduce to the Rayleigh-Jeans approximation 
$B_\nu(T_\text{dust})\propto\nu^2$.  A more accurate expression
can be obtained by approximating the Planck function as a power law 
$B_\nu(T_\text{dust}) \propto \nu^{\alpha_\text{Pl}}$, where 
$\alpha_\text{Pl} = \alpha_\text{Pl}(T_\text{dust}) \lesssim 2$ 
(and $=2$ in the Rayleigh-Jeans limit).  Given two frequencies, the spectral index $\alpha_\text{Pl}$ of the Planck function between $\nu_1$ and $\nu_2$ can be expressed as

\begin{equation}
\label{eqn:apl_def}
\alpha_\text{Pl} = \left\lvert \frac{\text{log}(B_{\nu_1}/B_{\nu_2})}{\text{log}(\nu_1/\nu_2)}\right\rvert
\end{equation}

\noindent Substituting the Taylor expansion of $B_\nu$ to second order yields the following approximation for the spectral index:

\begin{equation}
\label{eqn:apl_approx}
\alpha_\text{Pl} \approx 2 + \frac{\text{log}\left(\frac{2k_BT_\text{dust}-h\nu_1}{2k_BT_\text{dust}-h\nu_2}\right)}{\text{log}(\nu_1/\nu_2)}
\end{equation}

\noindent For our purposes, the spectral index of the Planck function can be approximated by the Rayleigh-Jeans solution with a correction that depends on the dust temperature $T_\text{dust}$ and observing frequencies used to determine the millimeter spectral index.  

Observationally, the flux density of debris disks at millimeter and centimeter 
wavelengths can be described by a simple 
power law $F_\nu \propto \nu^{\alpha_\text{mm}}$, where $\alpha_\text{mm} = \left\lvert \text{log}(F_{\nu_1}/F_{\nu_2})/\text{log}(\nu_1/\nu_2)\right\rvert$.  

Combining these relationships provides a simple expression for the 
slope of the grain size distribution, $q$ as a function of 
$\alpha_\text{mm}$, $\alpha_\text{Pl}$, and $\beta_s$:

\begin{equation}
\label{eqn:q}
q = \frac{\alpha_\text{mm}-\alpha_\text{Pl}}{\beta_s}+3
\end{equation}

\noindent Thus, by measuring the millimeter spectral index, $\alpha_\text{mm}$, and by inferring $\alpha_\text{Pl}$ from the dust temperature for each disk, 
we can determine the slope of the grain size distribution, $q$.

\subsection{Estimated $q$ Values for the Complete Sample}
\label{subsec:q_values}

\begin{figure}[ht]
\centerline{\psfig{file=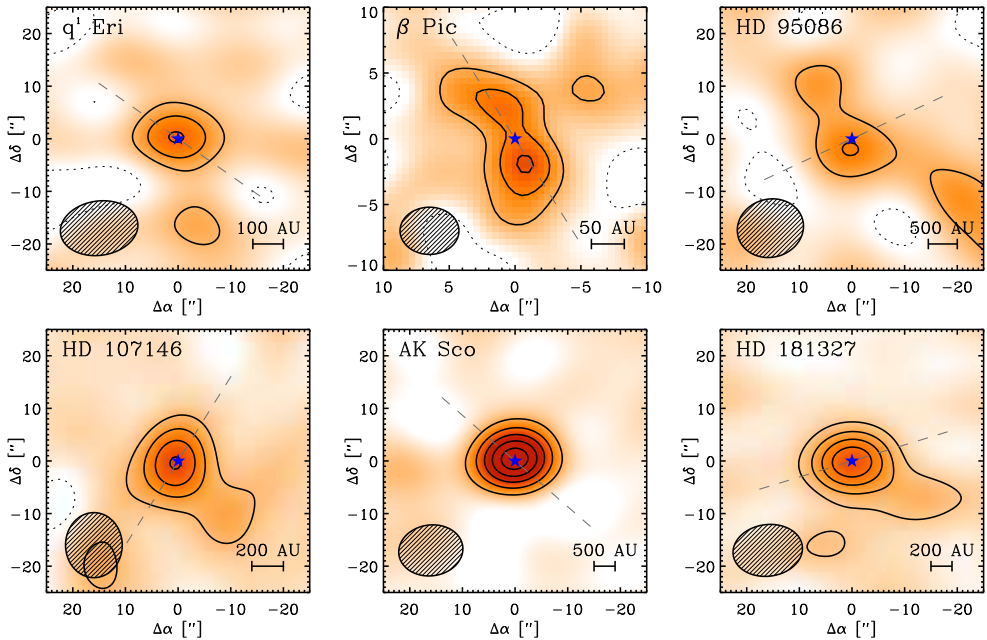,width=14cm,angle=0}}
\caption[]{\small Images of the 7 mm continuum emission from six debris disks 
detected in the ATCA sample. Contour levels are in steps of 2 $\times$ rms for all images, except AK Sco (lower center) where contours are in steps of 4 $\times$ rms (characteristic rms $\sim14$ $\mu$Jy). The ellipse in each lower left corner indicates the synthesized beam size.  The star symbol marks the position of the stellar photosphere.  The dashed gray line indicates the position angle of the disk determined from previous optical, near-infrared, or millimeter imaging and listed in Table~\ref{tab:sample}.}
\label{fig:fig2}
\end{figure}

We complement the VLA 9~mm observations of the seven debris disks with
ATCA 7~mm observations of eight additional debris disks (system 
characteristics listed in Table~\ref{tab:sample}) analyzed
in previous papers. Results for Fomalhaut were presented by \citet{ric12},
and results for q$^1$ Eri, $\beta$ Pic, HD 95086, HD 107146, and HD 181327 
were presented by \citet{ric15b}.  For HD 181327 and HD 95086, we have 
recalculated spectral indices given newly available ALMA measurements.
For q$^1$ Eri, we have recalculated the spectral index using the flux measurement at 870~$\mu$m from APEX \citep{lis08}, 
to be closer to the Rayleigh-Jeans regime for a more accurate determination of the spectral slope.
\cite{su15} suggest that previous flux measurements of HD 95086 are likely contaminated
by emission from the extragalactic background.  Any such background galaxies are included
within the ATCA beam and not resolved from the disk.  Given this situation, which may well
apply to other sources, we simply use the total flux density within the beam in the analysis.
 The $\epsilon$ Eridani debris disk, which was not detected at 7 mm with ATCA
\citep{mac15b}, provides an upper limit on the spectral index in combination
with resolved observations at 1.3 mm from the SMA.  
We have also included archival ATCA observations of AK Sco, a $\sim18$ Myr-old binary system with a massive ($\sim5-10 M_\text{Jup}$) circumbinary disk of gas and dust \citep{cze15}.  Given its age, there is some debate as to whether this system is a long-lived disk of primordial origin or a second-generation debris disk.  We choose to include it in the sample to examine the effects gas might have on collisional cascade properties (see Section~\ref{subsec:trends}).
We used the \texttt{uvfit} routine in Miriad to fit a point source model to 
the ATCA 7~mm AK Sco visibilities and obtained a total flux density of $F_\text{7mm} = 
430.0 \pm 18.2$~$\mu$Jy.  Figure~\ref{fig:fig2} shows images of the 7~mm continuum emission 
from the debris disks detected by ATCA (excluding Fomalhaut).

Table~\ref{tab:qvalues} lists a previously reported flux density of each disk in the sample at 
a given (sub)millimeter wavelength, and we use these values combined with 
the VLA 9~mm and ATCA 7~mm fluxes to determine a millimeter spectral index, 
$\alpha_\text{mm}$.  Eight of the disks in the sample have flux measurements 
from ALMA data available. For the remaining seven disks, we used flux 
measurements from the SMA, JCMT/SCUBA, and APEX as available.  
The reported uncertainties for these (sub)millimeter flux measurements include a $10\%$ uncertainty in the absolute flux density calibration added in quadrature to the statistical uncertainties.
Table~\ref{tab:qvalues} also presents the dust temperature ($T_\text{dust}$), the spectral index of the Planck function ($\alpha_\text{Pl}$), and the slope of the grain size distribution ($q$) derived using Equation~\ref{eqn:q}.  For all of the disks, the dust temperature was inferred by assuming radiative equilibrium with the central star: $T_\text{dust} = (L_*/16\pi\sigma R_\text{dust}^2)^{1/4}$.  Here, we used established stellar properties to determine $L_*$ and resolved (sub)millimeter imaging to estimate a characteristic radius for the dust, $R_\text{dust}$.  The uncertainties on $R_\text{dust}$ and $T_\text{dust}$ are conservatively estimated to be $20\%$ and $10\%$, respectively.  Given $T_\text{dust}$ and the two frequencies used to measure $\alpha_\text{mm}$, we determined $\alpha_\text{Pl}$ for each disk using the relationship derived in Equation~\ref{eqn:apl_approx}.  The final uncertainty on $q$ results from propagating the errors on $\alpha_\text{mm}$, $\alpha_\text{Pl}$, and $\beta_S$.

For the complete sample, $q$ values range from 2.84 (HD~141569) to 3.64 (HD~104860).  The weighted mean of 
these $q$ values is $\langle q \rangle = 3.36 \pm 0.02$.  This result 
is consistent with previous work by Ricci et al. (2015), which presented 
results for five of the debris disks included in this larger sample and obtained 
a weighted mean of $\langle q \rangle = 3.42 \pm 0.03$.

\subsection{Trends in $q$ with Stellar and Disk Properties}
\label{subsec:trends}

With this larger sample, we can not only determine a weighted mean value of 
$q$, but begin to look for trends with stellar and system characteristics.  
The stars in the sample span a wide range of ages ($5-4800$ Myr) and spectral 
types (A$-$M).  Table~\ref{tab:sample} lists the characteristics of the stars
in the full sample, including age, spectral type, luminosity, distance, 
and whether or not gas has been detected in the disk.  

We first consider trends with stellar properties, namely age and spectral type.  There does not seem to be any correlation between system age and grain size distribution.  For disks with estimated ages $<100$ Myr (49 Ceti, HD 15115, HD 61005, HD 141569, AU Mic, $\beta$ Pic, HD 95086, Ak Sco, and HD 181327), the weighted mean is $\langle q \rangle = 3.36\pm0.02$.  Disks with estimated ages $>100$ Myr (HD 377, HD 104860, q$^1$ Eri, $\epsilon$ Eri, HD 107146, and Fomalhaut) have a weighted mean of $\langle q \rangle = 3.35 \pm 0.02$, consistent with the other subsample within the uncertainties.  
A Kolmogorov-Smirnov (K-S) test gives a $85\%$ probability that these two subsamples are drawn from the same distribution.
However, we notice a tentative trend with spectral type.  If we arbitrarily choose to 
separate the sample into two groups by spectral type, 
then the weighted mean is $\langle q \rangle=3.40\pm0.02$ for stars with spectral types 
A$-$F (49 Ceti, HD 15115, HD 61005, HD104860, HD 141569, q$^1$ Eri, 
$\beta$ Pic, HD 95086, AK Sco, and HD 181327) and $\langle q \rangle = 3.30\pm0.03$ for 
spectral types G$-$M (HD 377, HD 61005, AU Mic, $\epsilon$~Eridani, and 
HD 107146).  Given the uncertainties, these two subsamples differ in $q$ by 
$\sim3\sigma$.  The clear outlier in the sample is HD 141569, with a $q$ value 
of $2.84\pm0.03$ (discussed in Section~\ref{subsec:stellar}).  If we exclude 
this source from the sample, the weighted mean for A$-$F stars is 
$\langle q \rangle=3.45\pm0.02$, different from the other subsample by $\sim5\sigma$.  
A Spearman's rank correlation measure of statistical dependence between two variables 
indicates that this trend is significant at 96\% confidence.  A K-S test gives a probability
of only $15\%$ that these two subsamples are drawn from the same distribution.
Given the small number statistics (there are noticeably fewer late type stars in the sample), 
we cannot draw any firm conclusions.  However, this is suggestive that stars 
with later spectral types may exhibit shallower grain size distributions. 
Interestingly, \cite{paw14} also note a trend of increasing $q$ values for more luminous stars, determined from mid- to far-infrared observations probing smaller grain sizes.

It is also plausible that characteristics of the disks themselves, regardless 
of stellar properties, might affect the grain size distribution.  Four of the 
disks in the sample have robust detections of gas: 49 Ceti, HD 141569, 
$\beta$ Pic, and AK Sco.  For these disks with gas, $\langle q \rangle=3.33\pm0.03$ including
HD 141569, and $\langle q \rangle=3.42\pm0.03$ omitting HD 141569, neither value differing 
from the mean of the full sample by $>3\sigma$.  The K-S probability for these
two subsamples is $92\%$. However, since there are only
four disks with detections of gas in the sample, observations of more debris 
disks are needed to address the effect of disk gas on grain size distribution.

\section{Discussion}
\label{sec:disc}

We have performed interferometric observations of a sample of seven debris disks 
at 9~mm with the VLA and supplemented them with observations of eight debris disks
observed with ATCA at 7~mm. By combining these long wavelength flux densities 
with previous (sub)millimeter measurements, we have determined a millimeter
spectral index and inferred the slope of the grain size distribution, $q$. The weighted mean of 
the $q$ values in the complete sample is $\langle q \rangle = 3.36 \pm 0.02$.   

We now compare these new results for the full sample to theoretical models of 
collisional cascades.  In particular, we consider the effects of
incorporating alternative velocity distributions 
(Section~\ref{subsec:collisional}), 
material strengths (Section~\ref{subsec:strength}), 
and radiation pressure blowout (Section~\ref{subsec:wavy}) on the resulting 
grain size distributions in debris disks.

\subsection{Comparison to Collisional Models}
\label{subsec:collisional}

The reference model for dust production in debris disks is the steady state 
catastrophic collisional cascade.  In this model, smaller `bullets' shatter 
larger `targets' through collisions.  Assuming conservation of mass, this 
shattering recipe leads to a power law size distribution of colliding bodies 
with radius $a$ within the disk, $n(a)\propto a^{-q}$.  A detailed description 
of the analytic treatment of collisional cascades can be found in \cite{pan05} 
and \cite{pan12}. 

The benchmark model of collisional cascades is presented by \cite{doh69}.  Using laboratory experiments, Dohnanyi formulated a model of collisions in the Asteroid Belt, where bodies dominated by material strength have an isotropic velocity dispersion and collisions occur between bodies of roughly the same size.  Given these assumptions, Dohnanyi obtains the classic result of $q = 7/2$ in steady-state.

If the assumption that the bodies participating in the collisional cascade have a single velocity dispersion regardless of size is relaxed, then steeper grain 
size distributions can be produced.  \cite{pan12} extend the \cite{doh69} formulation of collisional cascades by accounting for viscous stirring, dynamical friction, and collisional damping in addition to the mass conservation requirement already discussed.  For collisions between equal-sized, strength-dominated bodies, accounting for a size-dependent velocity distribution where velocity decreases with decreasing particle size yields $3.64 \leq q \leq 4$.  If, instead, the 
velocity \emph{increases} with decreasing particle size, then shallower size distributions are produced.  This scenario makes sense for particles close to the blowout size, where radiation pressure strongly affects the particles' velocities. 
Given a population of grains with velocity function $v \propto a^p$ and material strength parameterized by $Q_D^* \propto a^\gamma$, \cite{pan12} give a simple formulation for $q$ as a function of $\gamma$ and $p$: $q = \frac{21+\gamma-2p}{6+\gamma-2p}$.  For small grains ($a\lesssim1$~mm) strongly affected by radiation pressure, the collision velocities are expected to be proportional to the radiation-pressure induced eccentricities and $p\approx-1$.  For strength-dominated particles, $0~\geq~\gamma~>~-1/2$.  Given a representative value of $\gamma = -0.3$, this formulation yields $q = 2.95$.  For the somewhat larger (sub)millimeter particles probed by our observations, it is plausible to expect $q$ values somewhere between this estimation and Dohnanyi's $q=3.5$.

\cite{gas12} explored the evolution of collisional cascades in debris disks using numerical models.  They varied a total of 21 variables in their models describing the geometry of the system, the strength of the colliding bodies, and the outcome of collisions.  In steady-state, these calculations yield a range of $q \approx 3.64 - 4.33$.

A number of other numerical models incorporate more realistic physics in order to model the grain size distributions for specific systems.  \cite{loh12} use the ACE \cite[Analysis of Collisional Evolution, see e.g.][]{kri13} code, which takes into account grain material strength, mutual gravity, and the relative orientation of the orbits of colliding particles, to model the size distribution for HD 207129.  \cite{schu14} and \cite{schu15} use the same numerical code to model the size distributions of HIP 17439 and AU Mic, respectively.  For grain sizes between 10~$\mu$m and 1~mm, these models yield q values between 3.3 and 3.4 for all three systems.  Additionally, \cite{the08} suggest that the grain size distribution should depend on the degree of dynamical excitation of the dust-producing planetesimals.  In disks with parent bodies in low-eccentricity, low-inclination orbits, the resulting size distribution should be flatter than classical collisional cascade models.  \cite{paw15} use the ACE code to examine this prediction and obtain $q\sim3$ for grains of size $a = \frac{a_\text{bl}}{2e}$, where $a_\text{bl}$ is the blowout size and $e$ is the mean eccentricity of parent bodies.  For larger grains, $100<a<1000$~$\mu$m, the slope is closer to $\sim3.3-3.4$.

\begin{figure}[ht]
\centerline{\psfig{file=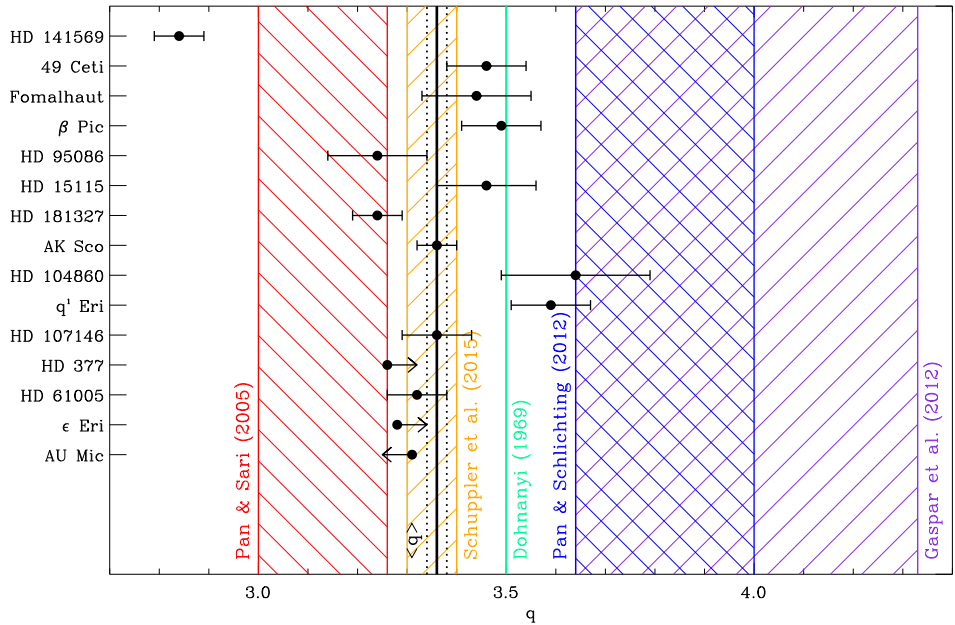,width=13.5cm,angle=0}}
\caption[]{\small Distribution of power law index $q$ values for the full sample of debris disks together with model predictions.  Black data points indicate the values for the individual disks in the sample, with the weighted mean and uncertainty, $q = 3.36 \pm 0.02$, shown by the solid and dotted black lines, respectively.  Stars have been ordered by luminosity from top to bottom (see Table~\ref{tab:sample} for specific luminosity values).  The solid lines and dashed regions indicate different model predictions: (red) `rubble pile' planetesimals not dominated by material strength \citep{pan05}, (orange) results of the ACE numerical model for AU Mic \citep{schu15}, (green) the classic \cite{doh69} result, (purple) numerical results of \cite{gas12}, and (blue) incorporating a size-dependent velocity distribution \citep{pan12}.
}
\label{fig:sample}
\end{figure}

Figure~\ref{fig:sample} shows the distribution of $q$ values from the fifteen debris disks in the sample along with the steady-state model predictions of \cite{doh69}, \cite{pan05}, \cite{schu15}, \cite{gas12}, and \cite{pan12}. The weighted mean of the sample, $\langle q \rangle = 3.36 \pm 0.02$, is comparable to the classic \cite{doh69} solution of $q = 3.5$, but still differs formally by $\sim7\sigma$.  Models of specific debris disk systems using the ACE numerical code \cite[e.g.][]{loh12,schu14,schu15} yield $q$ values between $\sim3.3-3.4$, lower than classical collisional cascade models and comparable to the weighted mean of our sample.
\cite{gas12} obtain some $q$ values from their models that are comparable to the observed mean value, but only by increasing the scaling of the strength regime, $S$, to $>> 10^{8}$ erg/g, in excess of the fiducial value of $3.5\times10^7$ erg/g \citep{benz99}.  All of the models from \cite{pan12} that incorporate a size-dependent velocity distribution where velocity decreases with decreasing particle size yield $q$ values greatly in excess of the observed $q$ values for all of the individual disks in our sample.  
However, the velocities in the grain size range of interest could instead increase with decreasing particle size due to radiation pressure.  As is shown above, such a velocity distribution could reproduce $q$ values comparable to our results.

We can also explore the effect of additional parameters on the grain size distribution, namely the strength of the colliding bodies and deviations from a strict power law introduced by a small-size cutoff due to radiation pressure.  This is done in the following two sections.

\subsection{Varying the Strength of Colliding Bodies}
\label{subsec:strength}

Models of collisions between rubble pile grains can predict $q$ values comparable to our results.  However, it is unclear if these parameters are realistic for
the colliding grains probed by our millimeter observations of debris disks.  The idea of collisions between strengthless rubble piles was proposed to explain observations of comets and Kuiper Belt Objects (KBOs), which are on the order of several kilometers in size \citep{asph96,jew02}.  Our observations probe millimeter to centimeter sized grains, which are typically assumed to be rocky and strength dominated.   Analytical and numerical calculations indicate that bodies do not become gravity-dominated until they are larger than $\sim 1$ km \citep{wya11}.  Furthermore, laboratory experiments show that the strength of small rocky particles does not vary significantly with particle size, scaling as $a^{-0.4}$ \citep{hous99}.  However, one could imagine that millimeter and centimeter sized grains might be more similar to loose conglomerates than rocky pebbles, and thus might exhibit strength scaling laws more similar to large rubble piles. In fact, measurements of the optical polarization of the AU Mic debris disk taken with \emph{HST/ACS} suggest that the grains in the disk are highly porous ($91-94\%$), more like `bird nests' \citep{gra07}.  

To produce the models discussed in Section~\ref{subsec:collisional}, we require that the bodies participating in the collisional cascade are strength-dominated.  If the colliding bodies are more like `rubble piles,' held together by gravity instead of material strength, we can replace the destruction criterion that collisions occur between bodies of the same size with the requirement that the kinetic energy of the bullet be equal to the gravitational energy of the target. \cite{pan05} derive a range of $q$ values $3.0 \leq q \leq 3.26$ for such a collisional population, lower than the steady-state collisional cascade predictions of \cite{doh69}.

\subsection{Wavy Distributions Produced by a Small-Size Cutoff}
\label{subsec:wavy}

Grains within a debris disk are continually subject to radiation pressure from the central star.  Large grains are less affected by this radiation pressure and remain in bound orbits, while smaller grains are placed in hyperbolic orbits.  A `blowout' size, $a_\text{bl}$, can be defined as the grain size for which the force due to radiation pressure ($F_\text{rad}$) is half the force from gravity ($F_\text{g}$) and $F_\text{rad}/F_\text{g} \gtrsim 0.5$.  For grains smaller than this blowout size, bound orbits are impossible and grains are removed from the disk.  

\cite{doh69} did not include a small-size cutoff in his theoretical framework to account for the removal of grains smaller than the blowout size.  Including such a cutoff superimposes waves or ripples on the predicted power law distribution.  The lack of grains smaller than $a_\text{bl}$, causes the equilibrium number of blowout-sized grains to be enhanced.  In turn, this results in an enhanced destruction rate of grains with sizes typically destroyed by blowout-sized grains.  Consequentially, the absence of these grains produces a higher equilibrium number of grains that would have been destroyed by them.  This ripple effect propagates upwards through the grain size distribution.  Previous numerical simulations have produced and discussed this wavy pattern \citep{camp99,the03,kriv06,wya11}.  The wavelength and amplitude of these waves depends strongly on the collisional velocities and properties of the colliding bodies. 

A realistic debris disk, however, is not expected to have a sharp cutoff at small grain sizes.  \cite{kriv06} note that dispersion in densities and fragmentation energies within an inhomogeneous grain population will likely weaken or smear any waves produced in the size distribution.  Additionally, including erosive or cratering collisions has a clear impact on the grain size distributions \citep{the07,ken16}, washing out ripples at grain sizes much above the blowout size.

Within the analysis of this paper, we have assumed that the size distribution of particles within a debris disk is a power law, $n(a)\propto a^{-q}$.  Given the wavy size distributions predicted by some numerical models, however, it is plausible that the power law size distribution we measure between millimeter and centimeter wavelengths does not reflect the full distribution from micron to kilometer sizes.  In order to explore the effect of ripples on the measured grain size distribution further, we have replicated with small modifications the numerical model presented by \cite{wya11} and implemented analytically by \cite{ken16}.  The details of the model are described in Section~\ref{subsubsec:wavy_model} and the results and implications are 
presented in Section~\ref{subsubsec:q_waves}.  This simple model provides a useful illustration of how possible modulations in the grain size distribution would be reflected in the observed opacity spectrum.  However, the results should not be over interpreted, given the range of effects that may damp such modulations in more realistic systems.

\subsubsection{A Steady State Model with a Small-Size Cutoff}
\label{subsubsec:wavy_model}

We define a population of planetesimals divided into $N$ bins spaced logarithmically in size, where the mass and size of the $k$th bin are $m_k$ and $a_k$, respectively.  The largest bin is defined to be $k=1$ and size decreases with increasing $k$ such that $a_{k+1}/a_k = 1-\delta$, with $\delta = 0.01$.  The size of the smallest bin is defined as the blowout size, $a_\text{bl}$, and the size of the largest bin is fixed at $10^4$ m.  Like \cite{wya11}, we assume a steady state where the mass loss rate per logarithmic bin is constant and the mass in each bin is defined as $m_k = C/R_k^c$, where $C$ is some arbitrary constant and $R_k^c$ is the collision rate in bin $k$.  The discrete form of the collision rate can be expressed as

\begin{equation}
\label{eqn:rkc}
R_k^c = \sum_{i=1}^{i_{ck}} \frac{3m_i}{2\rho\pi a_i^3}(a_k+a_i)^2P_{ik}\text{,}
\end{equation}

\noindent where $\rho$ is the particle density, $P_{ik}$ is the intrinsic collision probability between particle $i$ and $k$ defined as $\pi v_\text{rel}/V$, $v_\text{rel}$ is the relative collision velocity, and $V$ is the total volume through which the planetesimals are moving.  The smallest impactors that can catastrophically destroy particles of size $a_k$ have size $X_ca_k$, where $X_c = (2Q_D^*/v_\text{rel}^2)^{1/3}$ and $Q_D^*$ is the collision energy required to eject half the mass from a pair of colliding bodies.

Using this formalism, the mass in each bin can be solved for analytically, beginning with the smallest bin, $k=N$.  Since there are no smaller impactors available, only bodies of size $a_N$ are involved in the collisional cascade.  Thus, $R_K^c$ can be solved for simply:

\begin{equation}
\label{eqn:rnc}
R_N^c = \left(\frac{6P_{NN}}{\rho\pi a_N}\right)\times m_N = A_Nm_N
\end{equation}

\noindent The mass in bin $N$ is then $\sqrt{C/A_N}$.  Next, we can consider bin $N-1$.  In this bin, the summation in Equation~\ref{eqn:rkc} has two terms:

\begin{equation}
\label{eqn:two_terms}
R_{k}^c = A_{k+1}m_{k+1}+ A_{k}m_{k}
\end{equation}

\noindent Since we already determined the mass in bin $N$, we know the first term $A_{k+1}m_{k+1} = B_k$.  We can then simplify Equation~\ref{eqn:two_terms} to $R_k^c = B_k + m_kA_k$.  The mass in bin $N-1$ is then found by solving a quadratic equation:

\begin{equation}
\label{eqn:quadratic}
m_k^2A_k + m_kB_k - C = 0
\end{equation}

\noindent  Moving up to bins of larger size, $R_k^c$ always contains two terms, a sum over all collisions with smaller particles and $m_kA_k$ for the current bin.  Thus, the mass in each bin is found simply by solving Equation~\ref{eqn:quadratic} for every bin.  In this formulation, the shape of the steady-state distribution is independent of the total mass.

Once we have determined the steady-state number of grains in each bin, we calculate grain opacities to determine whether the resulting wavy number distribution translates to an observable effect.  We assume the same grain compositions as \cite{ric10a} that contain 7\% silicates, 21\% carbon, and 42\% water ice by volume, and have 30\% porosity.  The details of these assumptions are not important to demonstrate the resulting effect on the opacities.  For each size bin, we use the Mie scattering code implemented by \cite{dull04} for use in RADMC, a code for dust continuum radiative transfer.  The output of this code is an opacity spectrum for each individual grain size in our model.  To determine the ensemble opacity at each wavelength, we then calculate a mass-weighted average over all grain sizes given the steady-state size distribution.

The final result of this procedure is the ensemble opacity of a given population of grains as a function of wavelength.  For our observations, we measure flux density at two wavelengths in order to determine the millimeter spectral index, $\alpha_\text{mm}$.  
If the dust opacity is a simple power law, $\nu^\beta$, we can infer $\beta = \alpha_\text{mm}-2$ from our observations so long as the disk emission is optically thin and in the Rayleigh-Jeans regime.  Given the final ensemble opacity spectrum from our models, we can determine the power law index, $\beta$, between any two wavelengths we might observe at.  By computing $\beta$ for a range of models and wavelengths corresponding to the observations presented in this study, we can examine the scatter in $\beta$ (and thus $q$) we might expect due to waves produced by a small-size cutoff in the grain size distribution.

\subsubsection{Variations in Spectral Indices Due to Waves in Grain Size Distributions}
\label{subsubsec:q_waves}

To determine the expected scatter in the observed millimeter-centimeter spectral index from waves superimposed on the power law grain size distribution, we considered three free parameters in our model: $a_\text{bl}$, $v_\text{rel}$, and $Q_D^*$.  For debris disks, $v_\text{rel} \gtrsim 1$ km s$^{-1}$ \citep{kri07}; we calculated models for a range of 1 to 6~km~s$^{-1}$.  We considered both constant $Q_D^* = Q_s$ and a power law dependence on grain size, $Q_D^* = Q_sr^{\beta_s}$.  For rocky objects dominated by material strength, $Q_s = 6\times10^3$~J kg$^{-1}$ and $\beta_s = -0.40$ \citep{benz99}.  The blowout size, $a_\text{bl}$ is expected to vary between $\sim 1 - 10$ $\mu$m for the disks in the sample.

\begin{figure}[ht]
\begin{minipage}[h]{0.5\textwidth}
  \begin{center}
       \includegraphics[scale=0.65]{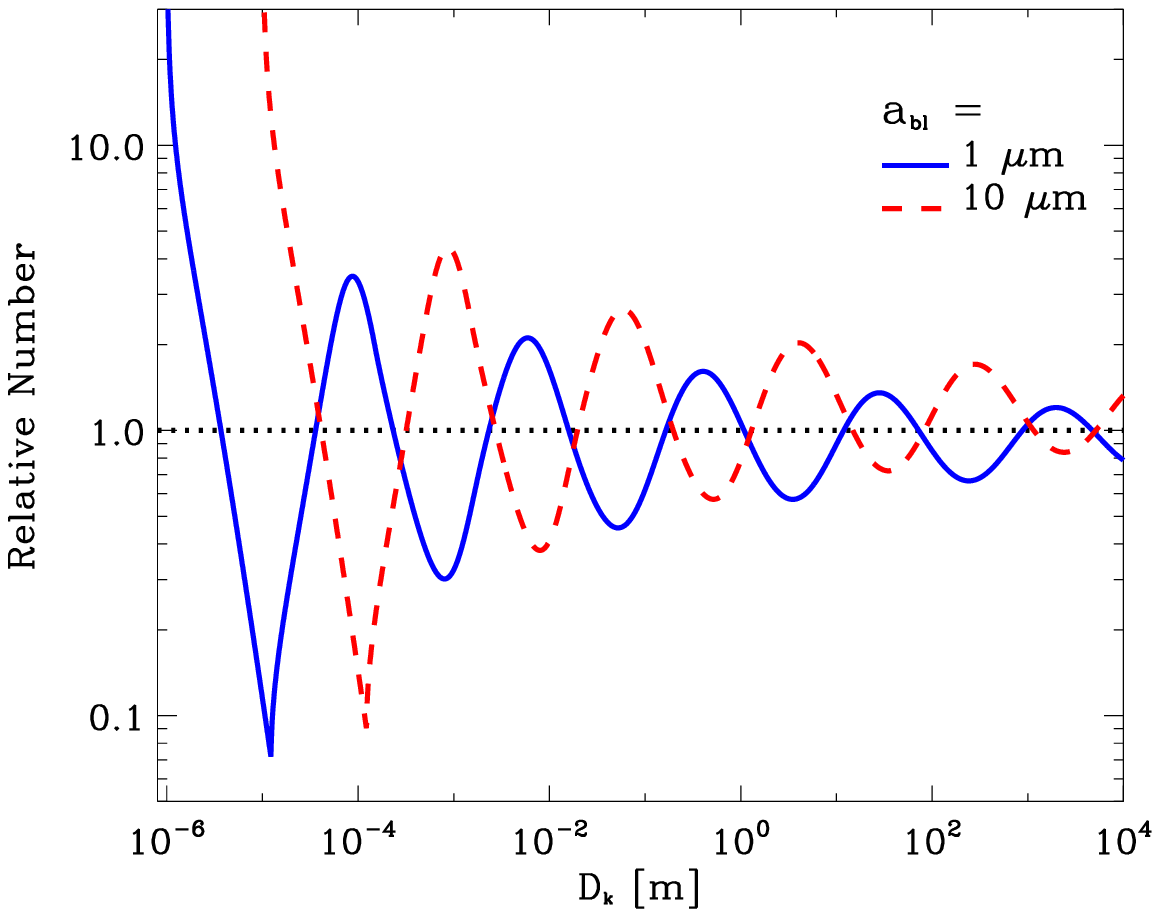}
  \end{center}
 \end{minipage}
 \begin{minipage}[h]{0.5\textwidth}
  \begin{center}
       \includegraphics[scale=0.65]{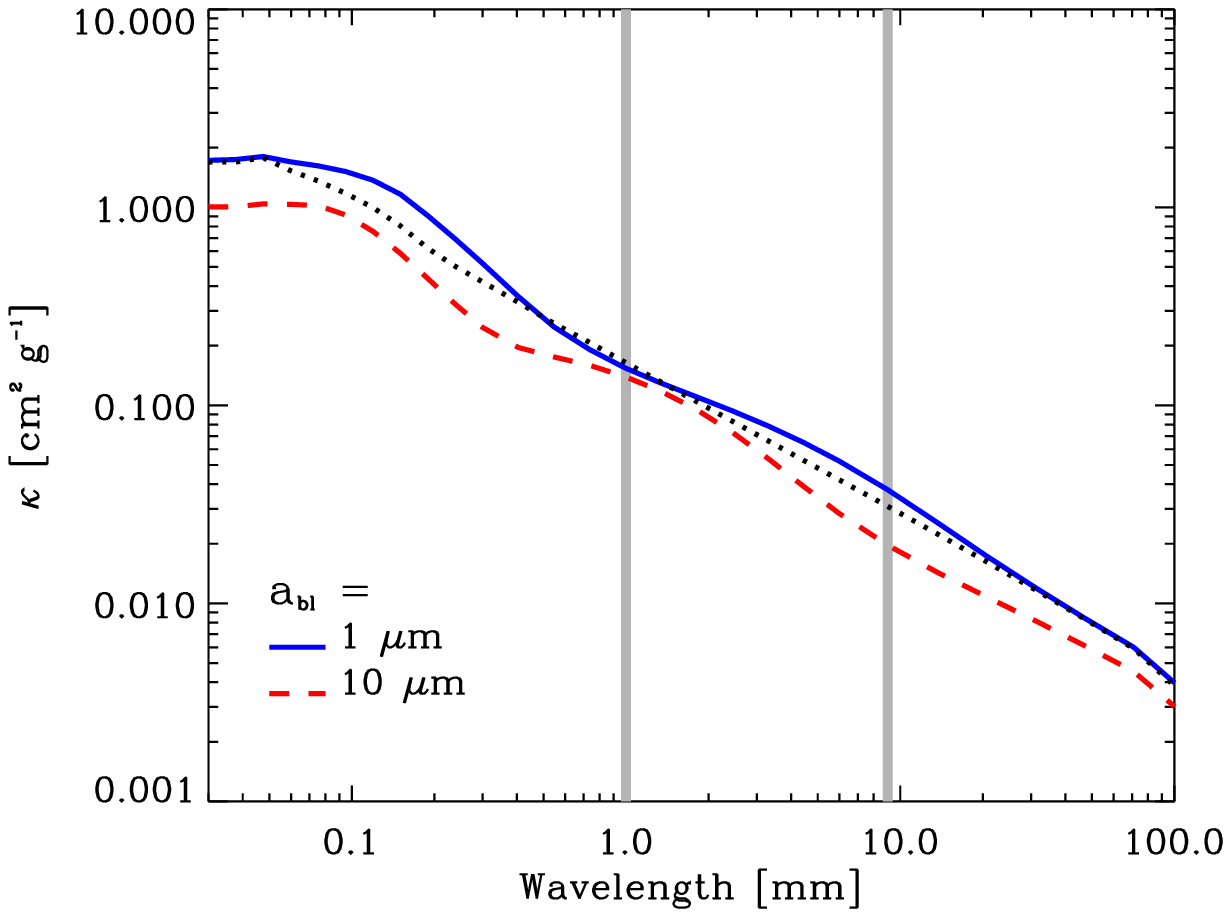}
  \end{center}
 \end{minipage}
 \caption{\small \emph{(left)} Comparison of the relative size distribution, $n(a)/n(a)_\text{pl}$, for a steady-state collisional cascade with a blowout size $a_\text{bl}$ of 1 $\mu$m (blue solid line) and 10 $\mu$m (red dashed line).  For both models, we have divided the analytic number distribution, $n(a)$, by the expectation for a power law distribution, $n(a)_\text{pl}\propto a^{-3.5}$, to show the waves introduced by the small-size cutoff more clearly.  The expected power law size distribution, $n(a)_\text{pl}$, is indicated by the dotted black line.  For all models, we assume $Q_D^* = 6\times10^3$ J kg$^{-1}$ and $v_\text{rel} = 5$ km s$^{-1}$.  
 \emph{(right)} The resulting ensemble opacity as a function of wavelength for a 1 $\mu$m blowout size (blue solid line) and a 10~$\mu$m blowout size (red dashed line).  Again, the dotted black line indicates the expected ensemble opacity for the power law size distribution with no waves.  The gray lines indicate the wavelengths of our (sub)millimeter and VLA observations used to determine the millimeter spectral index, $\alpha_\text{mm}$.
 }
\label{fig:models}
\end{figure}

Figure~\ref{fig:models} shows the resulting size distribution (left panel) and ensemble 
opacity as a function of wavelength (right panel) for two example models with 
blowout size, $a_\text{bl}$, of 1~$\mu$m and 10~$\mu$m.  For both models, 
we fixed $v_\text{rel}=5$ km s$^{-1}$ and $Q_D^*= 6 \times 10^3$ J kg$^{-1}$.  
The waves produced in the size distribution by the changing blowout size are 
seen as muted features in the dust opacity curves.  
For these two models, blowout sizes of 1 and 10 $\mu$m produce 
dust opacity power law indices of $\beta = 0.63$  and $0.88$, respectively.  
For lower relative velocities ($v_\text{rel}\sim1$~km~s$^{-1}$) the waves 
are damped at long millimeter and centimeter wavelengths and the change in 
 $\beta$ is negligible.  For the full range of free parameters we explored, the resulting values of $\beta$ ranged from 0.5 to 1.1, comparable to the spread in our complete sample of debris disks, $-0.20 \lesssim \beta \lesssim 1.15$.
This is not a definitive explanation for the trend we see in spectral index as a function of stellar type, but it suggests that the waves produced by introducing a cutoff at small grain sizes could introduce observable scatter in the 
values of $\beta$, and thus $q$, similar to that seen in the measured spectral indices.
It is also possible that grain porosity could change with particle size and create damped resonances as a function of wavelength that might produce structure in the opacity spectrum and result in low values of $\beta$.
Recent work suggests that K$-$M (and possibly G) stars do not exhibit a blowout limit for plausible dust compositions \cite[e.g.][]{reid11,vit12,schu15}.  The lack of a sharp size cutoff for such stars could further damp any waves in the grain size distribution, possibly contributing to the inferred shallower size distribution for these late-type stars.

\subsection{Stellar Emission Components}
\label{subsec:stellar}

A potential source of bias is our implicit assumption that all of the emission
we detect at long wavelengths comes from the dusty debris.  
If the measured flux densities at long wavelengths include emission from any 
additional mechanisms, then we will 
underestimate the true millimeter spectral index and thus the size distribution
power law index, $q$.  As discussed in the appendix, the AU Mic system exhibits 
significant emission from stellar activity at both 1.3 and 9 mm that can 
be explained by models of a hot stellar corona or chromosphere.  Given this stellar activity,
the presence of an asteroid belt \citep{mac13,schu15} is no longer needed to explain
the previously reported compact millimeter emission.  In this 
system, we were able to distinguish the excess stellar emission from the disk 
and subtract it using a time-dependent model. Only three other systems are
well enough resolved by the centimeter observations to address this issue,  
$\beta$ Pic, $\epsilon$ Eridani, and Fomalhaut. Indeed, while the $\epsilon$ Eridani debris disk is not detected at centimeter wavelengths, the central star exhibits excess emission attributable to a hot chromosphere \citep{mac15b}.  The Fomalhaut system shows a central peak
at 7~mm distinct from the cold, outer debris belt whose origin is unclear
\citep{ric12}.  In the eleven unresolved systems in the sample, we cannot 
separate disk dust emission from any stellar contamination, if present. 

The clear outlier in the sample is HD 141569, with a $q$ value of 
$2.84\pm0.03$, well-below the weighted mean of our complete sample and 
the classic Dohnanyi prediction.  In order to increase this $q$-value to 
3.50, as predicted by \cite{doh69}, more than $90\%$ of the emission measured 
at 9~mm would have to come from contamination from a stellar component, vastly in excess of the expected photospheric flux at these long wavelengths.  However, a contribution from an active stellar chromosphere or corona cannot be ruled out.  While this system has been imaged with ALMA at 870 $\mu$m, it remains unresolved (White et al. in prep), so we have no constraints on the millimeter or centimeter stellar emission.  The HD 141569 disk is the youngest source in the sample with an age of $\sim5$ Myr and contains a significant amount of gas.  Furthermore, HD 141569 has two nearby M dwarf companions \citep{wei00} with their own associated radio emission (see Figure~\ref{fig:fig1}).  It is possible that one or 
all of these system characteristics contributes to the low $q$ value measured, 
or that the central region of the disk is optically thick at these wavelengths.
For example, gas drag and other transport mechanisms (Poynting-Robertson or stellar wind drag) would tend to flatten the size distribution of affected grains.  \cite{wya11} show that the resulting $q$ value is $\alpha_r-1$, where $\alpha_r$ is the slope of the redistribution function or the mass distribution of fragments produced in collisions.  Given $\alpha_r=4$, the expected $q$ value for a system with gas drag would be $\sim3$, much flatter than the classic Dohnanyi prediction of $q=3.5$ and more comparable to the result for HD 141569.

In order to conclusively determine if the flux measurement of HD 141569 or
any of the other unresolved disks in the sample are contaminated by coronal 
or chromospheric emission, observations at millimeter and centimeter 
wavelengths with higher angular resolution are needed.  A growing number of nearby solar type stars, including $\alpha$ Cen A and B \citep{lis15} and $\epsilon$ Eridani \citep{mac15b}, have all been seen to exhibit excess emission at long wavelengths attributable to a hot chromosphere.  However, except for AU Mic, none of the other debris disks in our sample that have been resolved with ALMA show any evidence for an additional strong stellar emission component.

\section{Conclusions}
\label{sec:concl}

We present new VLA observations at 9~mm of a sample of seven debris disks. Using the best available flux measurements at (sub)millimeter wavelengths, we place tight constraints on the millimeter/centimeter spectral indices. We combine these with archival ATCA observations at 7~mm of an additional eight debris disks building on the work of \cite{ric15} and \cite{ric12} to infer the dust grain size distribution power law index, $q$.    

\begin{enumerate}

\item For the full sample, the weighted mean for the slope of the power law grain distribution is $\langle q \rangle=3.36\pm0.02$, 
with a range of 2.84 to 3.64.  This result is closest to the prediction of $q = 3.50$ in the classical model 
of a collisional cascade presented by \cite{doh69} and to recent numerical results by \cite{loh12}, \cite{schu14}, and \cite{schu15} for specific debris disk systems.  The models of \cite{pan12} that incorporate size-dependent velocity distributions where velocity decreases with decreasing particle size produce significantly steeper size distributions.  Numerical models by \cite{gas12} yield some $q$ values consistent with these results, but only by increasing the scaling of the strength curve in excess of fiducial values.  Shallower size distributions can be produced by models that allow for a velocity distribution where velocity increases with decreasing particle size or models that consider colliding bodies not dominated by material strength. 

\item Although limited by small number statistics, the observations suggest a trend in $q$ as a function of stellar type.  The weighted mean for stars with spectral types A$-$F is $3.45\pm0.02$ and $3.30\pm0.03$ for spectral types G$-$M.    
We see no evidence for trends in $q$ as a function of system age or gas abundance in the disk.  

\item Introducing a cutoff in the grain size distribution at small sizes due 
to radiation pressure can superimpose waves on the power law grain distribution.  
We examine a range of analytic models varying the blowout size ($a_\text{bl}$),
the relative collision velocity ($v_\text{rel}$), and the collision energy 
($Q_D^*$).  Changing the blowout size from 1 $\mu$m (typical for K and M stars)
to 10 $\mu$m (for A and B stars) produces measurable waves in the grain size 
distribution and the resulting dust opacity.  For a reasonable range of 
parameter values, the waves produced vary the inferred dust opacity power law index $\beta$ between 0.5 and 1.1,
a spread in values comparable to the scatter in the observations.  In realistic debris disk systems, however, inhomogeneities in densities and fragmentation energies likely weaken or smear these modulations.

\item The VLA observations of the AU Mic system show significant and variable
emission from stellar activity at centimeter wavelengths on timescales from 
minutes to months.  Given this new evidence, the asteroid belt posited by 
\cite{mac13} is not needed to explain any aspects of the observed millimeter 
emission.  High angular resolution at centimeter wavelengths allows us to distinguish this 
stellar emission from the dust disk emission.  However, the vast majority of 
debris disks in the sample have not been resolved at centimeter wavelengths.  
Given this, we are unable to separate disk dust emission from any stellar 
contamination (or extragalactic background contamination), if present, for 
these systems.  Higher resolution observations at millimeter and centimeter 
wavelengths are needed to place better constraints on the contribution of 
stellar activity to the total flux measurements.

\end{enumerate}

These VLA and ATCA observations provide the longest wavelength flux measurements of these fifteen debris disk systems to date.  But, observations at centimeter wavelengths of new systems are needed to continue to grow the sample and place better constraints on any trends in the grain size distribution with stellar or disk properties.

\acknowledgements
M.A.M acknowledges support from a National Science Foundation Graduate Research Fellowship (DGE1144152).  The National Radio Astronomy Observatory is a facility of the National Science Foundation operated under cooperative agreement by Associated Universities, Inc.  The Australia Telescope Compact Array is part of the Australia Telescope National Facility which is funded by the Australian Government for operation as a National Facility managed by CSIRO.  We thank Margaret Pan, Hilke Schlichting, and Scott Kenyon for helpful conversations.  We thank Meredith Hughes, Kate Su, David Rodriguez, Sebasti{\'a}n Marino, Aaron Boley, and Jacob White for providing data prior to publication.  We also thank the anonymous referee for a careful and thoughtful review.

\bibliography{References}

\begin{appendices}
\section{Radio Light Curves of AU Mic}
\label{appendix}
\end{appendices}

AU Mic is an active M dwarf star that is known to exhibit radio-wave bursts. In quiescence, previous observations placed upper limits on the flux at radio wavelengths, $<~120$~$\mu$Jy at 3.6~cm \citep{white94,leto00}.  Recent ALMA observations of the system at 1.3~mm revealed a compact central emission peak with a flux of $\sim320$~$\mu$Jy in addition to the continuum dust emission from the debris belt and greatly in excess of the expected photospheric flux at this long wavelength.  \cite{cra13} model this excess emission as arising from a hot stellar corona.  \cite{schu15} suggest that chromospheric emission from the star could also contribute at millimeter and radio wavelengths.

These new VLA observations of AU Mic offer a unique glimpse at the variations in radio emission from the star on minute to month-long timescales. Observations were taken at 9~mm (Ka band) in the DnC configuration on May 9 and 11, 2013 and in the C configuration on June 21, 2013.  Each of the three scheduling blocks (SBs) was a total of 105 minutes in length.  Observations of AU Mic were interleaved with the gain calibrator, J2101-2933, in a 4 minute cycle, with 3 minutes on-source.  

\begin{figure}[ht]
\begin{minipage}[h]{0.32\textwidth}
  \begin{center}
       \includegraphics[scale=0.52]{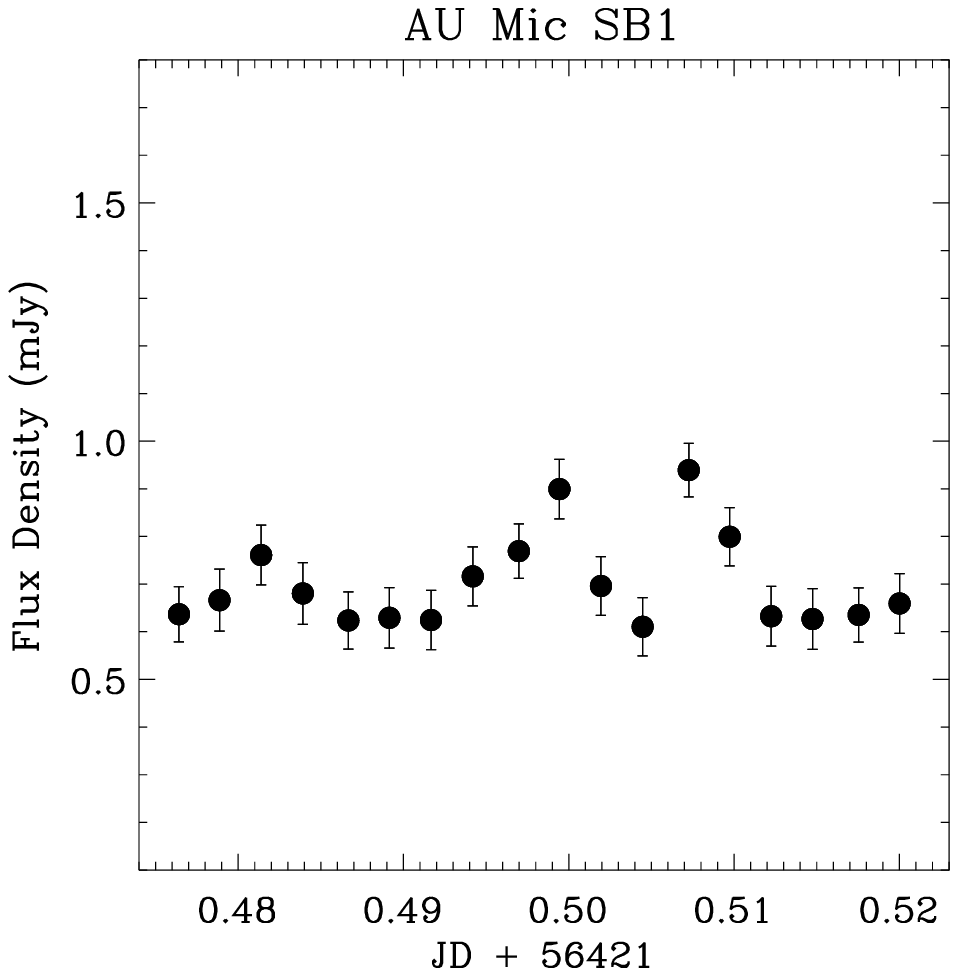}
  \end{center}
 \end{minipage}
 \begin{minipage}[h]{0.32\textwidth}
  \begin{center}
       \includegraphics[scale=0.52]{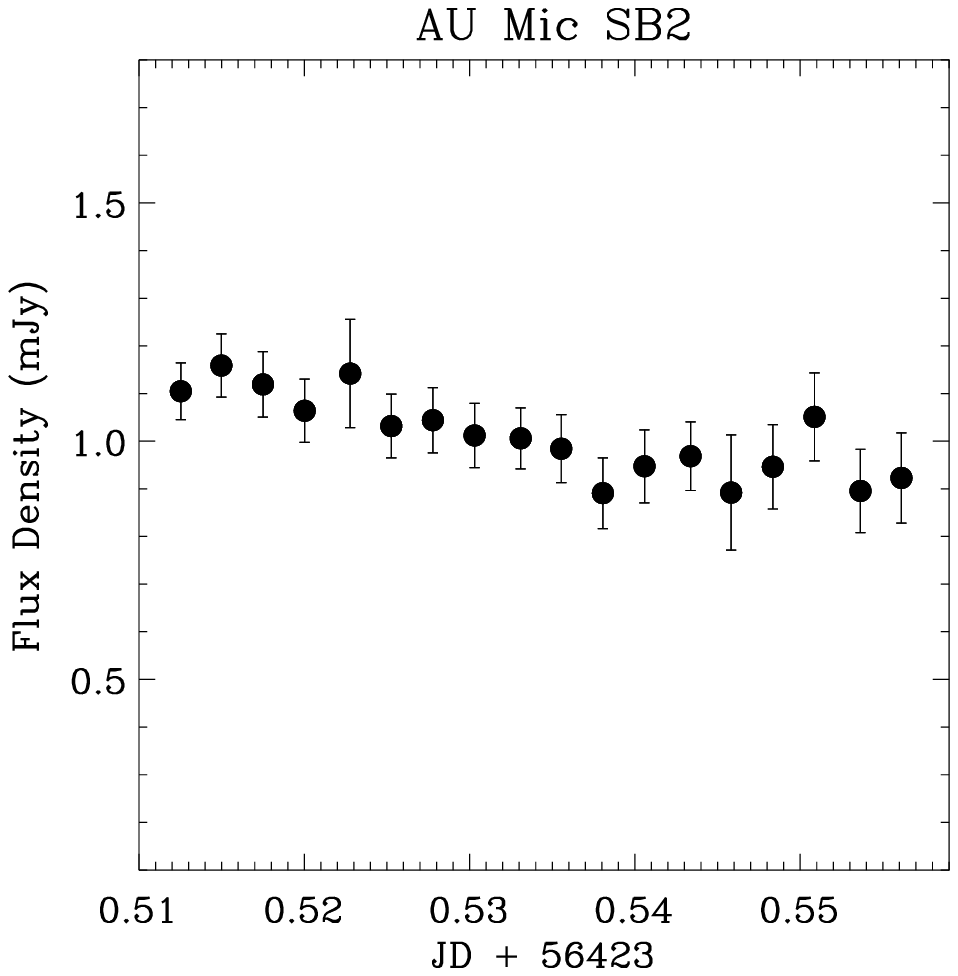}
  \end{center}
 \end{minipage}
  \begin{minipage}[h]{0.32\textwidth}
  \begin{center}
       \includegraphics[scale=0.52]{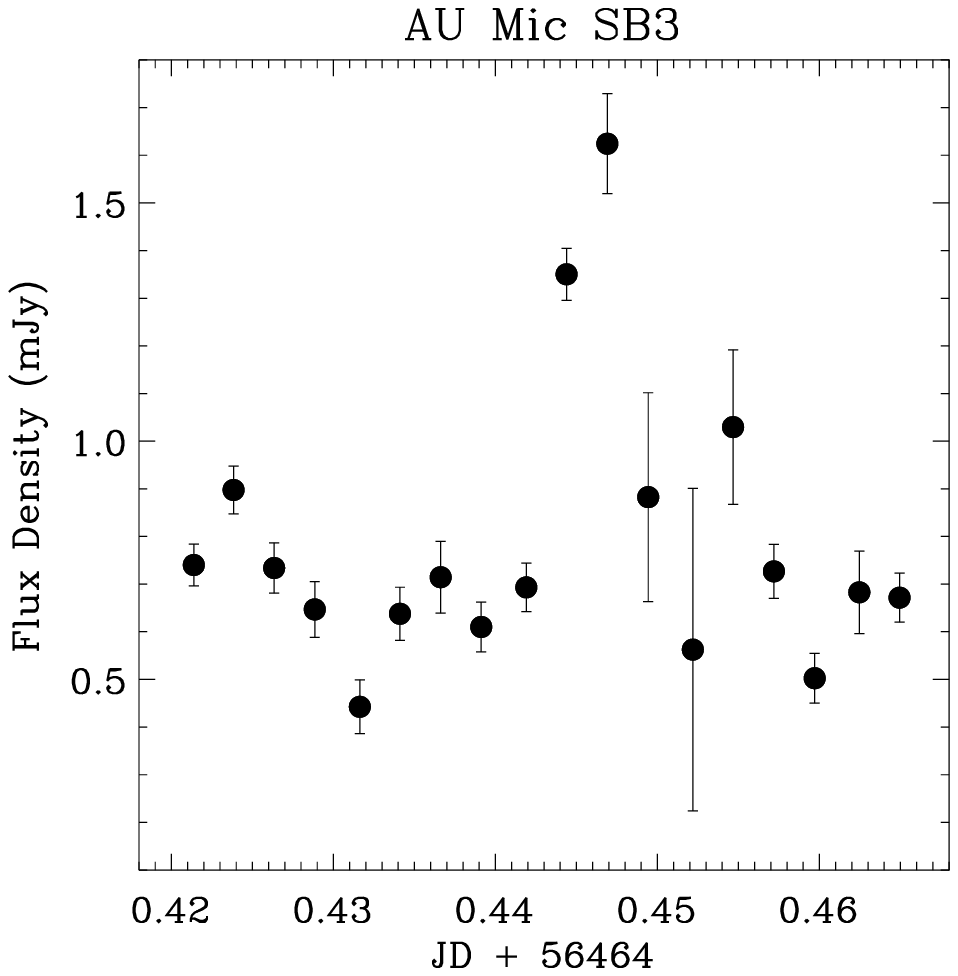}
    \end{center}
  \end{minipage}
\caption{\small Stellar light curves for AU Mic showing the variation in flux density over the course of our three observations.  The flux density was determined for each 3 minute integration using the CASA task \texttt{uvmodelfit} to fit a point source model to the visibilities.
}
\label{fig:aumic}
\end{figure}

Figure~\ref{fig:aumic} shows the stellar light curves for each of the three SBs.  To determine the flux density, we used the CASA task \texttt{uvmodelfit} to fit a point source model to the visibilities from each 3 minute on-source integration.  There is strong radio emission from the star present in all 3 SBs with a flux of $0.5 - 1.5$~mJy.  During the third SB, the star appears to have flared strongly, increasing in brightness by $100-150\%$.

\citet{cra13} simulated the time-steady coronal emission of AU Mic in both
X-ray and millimeter bands, but they did not account for time variability.
In order to begin to estimate the relative enhancements in the coronal
emission during flare times, we re-ran the \citet{cra13} model for
a range of imposed variations in the coronal heating.
We assumed that a fractional filling factor, $f$, of the coronal volume
was occupied by flare-like plasma, and that the photospheric driving
velocity, $v_{\perp}$, of MHD turbulence was enhanced by a dimensionless
factor, $\xi$, in that region.
Thus, for a range of input parameters ($f \leq 1$, $\xi \geq 1$) we
simulated both the total X-ray luminosity, $L_{\rm X}$, and the
thermal free-free emission, $I_{\nu}$, at the observed VLA wavelength
of 9 mm.  The goal was to find a combination of plausible $f$ and $\xi$ values
that would produce a relative enhancement in $L_{\rm X}$ of order
30\% to 60\%
\cite[to match the flare amplitudes reported by][]{smi05,mit05,sch10}
simultaneously with an increase in the 9~mm continuum of order
$100-150\%$ (see Figure~\ref{fig:aumic}).

To match the observed ranges of flux density enhancement, we found that the
filling factor $f$ must be larger than 0.48, and the boundary velocity
enhancement factor $\xi$ must be within the range of $4.1-8.2$.
These values of $\xi$ correspond to an increase in the total nonthermal
energy input of order $20-70$ in the flare regions.
The maximum enhanced temperature in this subset of models was found
to be about 13 MK.
This is reminiscent of the hottest of the three temperature
components (3.37, 7.78, and 17.3 MK) inferred from {\em Chandra}
spectra of AU~Mic by \citet{sch10}.
The quiescent model of \citet{cra13} successfully predicted temperatures
between the two lower {\em Chandra} values, but the largest value
appeared to be explainable only by flare activity.

A better understanding of cool-star flare activity will depend on
broader observational coverage in both time and wavelength.
{\em Kepler} has opened up new vistas of study of how properties
of white-light ``superflares'' on active stars may relate to the
more familiar case of the Sun \citep{mae15}.
However, recent multi-wavelength surveys of M and K dwarf exoplanet
host stars \cite[e.g.,][]{fra15} are also revealing new information
about weak flares that do not manifest at visible wavelengths.
For the better-resolved case of the Sun, recent observations and
simulations show that nearly {\em all} coronal heating
activity may take the form of a distribution of flare events with
a range of strengths and timescales \citep{fle15}.
Even models that invoke MHD waves and turbulence appear to require
the spontaneous production of nanoflare-like magnetic reconnection
events as a final product of the cascade from large to small eddies
\cite[e.g.,][]{osm11,vell15}.

 \begin{deluxetable}{l c c c c c c c c}
\tablecolumns{9}
\tabcolsep0.1in\footnotesize
\tabletypesize{\footnotesize}
\tablewidth{0pt}
\tablecaption{Debris Disk Sample Characteristics}
\tablehead{
\colhead{Source} & 
\colhead{$\alpha$ (J2000)} & 
\colhead{$\delta$ (J2000)} & 
\colhead{Spectral} & 
\colhead{L$_*$} &
\colhead{d$^\text{a}$} &
\colhead{Age} &
\colhead{PA$^\text{b}$} &
\colhead{Disk} \\
\colhead{} &
\colhead{} & 
\colhead{} & 
\colhead{Type} &
\colhead{[$L_\odot$]} & 
\colhead{[pc]} &
\colhead{[Myr]} &
\colhead{[$\degr$]} &
\colhead{gas?}
}
\startdata
HD 377 &  00 08 25.8 & $+06$ 37 00.5 & G2 & 1.0 & 30 & 150 & 47 & N \\
49 Ceti & 01 34 37.8 & $-15$ 40 34.9 & A1 & 20. & 59 & 40 & 101 & Y \\
HD 15115 & 02 26 16.3 & $+06$ 17 33.1 & F2 & 3.3 & 45 & 21 & 279 & N \\
HD 61005 & 07 35 47.5 & $-32$ 12 14.0 & G8 & 0.5 & 35 & 40 & 70 & N  \\
HD 104860 & 12 04 33.7 & $+66$ 20 11.7 & F8 & 1.4 & 48 & 140 & 1 & N \\
HD 141569 & 15 49 57.8 & $-03$ 55 16.2 & B9.5 & 21. & 116 & 5 & 356 & Y \\
AU Mic & 20 45 09.8 & $-31$ 20 31.8 & M1 & 0.1 & 10 & 21 & 128 & N \\
\\
\hline
\\
q$^1$ Eri &01 42 29.3 & $-53$ 44 27.0 & F9 & 1.2 & 17 & 4800 & 55 & N  \\
$\epsilon$ Eri & 03 32 54.9 & $-09$ 27 29.4 & K2 & 0.3 & 3 & $400-800$ & 0 & N \\
$\beta$ Pic & 05 47 17.1 & $-51$ 03 59.4 & A6 & 8.7 & 19 & 21 & 32 & Y  \\
HD 95086 & 10 57 03.0 & $-68$ 40 02.5 & A8 & 8.6 & 90 & 17 & 15 & N  \\
HD 107146 & 12 19 06.5 & $+16$ 32 53.9 & G2 & 1.0 & 29 & $80-200$ & 148 & N  \\
AK Sco & 16 54 44.8 & $-36$ 53 18.6 & F5 & 3.0 & 142 & 18 & 49 & Y \\
HD 181327 & 19 22 58.9 & $-54$ 32 17.0 & F6 & 3.3 & 51 & 12 & 107 & N  \\
Fomalhaut & 22 57 39.0 & $-29$ 37 20.1 & A4 & 16. & 7 & 440 &156 & N \\
\enddata
\tablecomments{$^a$ All distances measured by \cite{vanL07}\\
$^b$ Position angle measured east of north to the disk major axis\\
References for stellar and disk properties: HD 377, \cite{geer12}, \cite{cho15}; 49 Ceti, \cite{tor08}, \cite{hug08}; HD 15115, \cite{bin14}, \cite{kal07}; HD 61005, \cite{des11}, \cite{hines07}; HD 104860, \cite{stee15}; HD 141569, \cite{wei00}, White et al. (in prep.); AU Mic, \cite{bin14}, \cite{mac13};  q$^1$ Eri, \cite{but06}, \cite{lis08}, \cite{lis10}; $\epsilon$ Eri, \cite{mam08}, \cite{gre14}; $\beta$ Pic, \cite{bin14}, \cite{den14}, \cite{heap00}; HD 95086, \cite{mes13}, \cite{su15}; HD 107146, \cite{wic03}, \cite{ard04}; AK Sco, \cite{pec12}, \cite{cze15}; HD 181327, \cite{nor04}, \cite{sch06}; Fomalhaut, \cite{mam12}, \cite{kal05}
}
\label{tab:sample}
\end{deluxetable}

\begin{deluxetable}{l c c c c c c c c}
\tablecolumns{8}
\tabcolsep0.06in\footnotesize
\tabletypesize{\footnotesize}
\tablewidth{0pt}
\tablecaption{VLA Observations}
\tablehead{
\colhead{Source} & 
\colhead{Obs.} & 
\colhead{Config.} & 
\colhead{Ant.} &
\colhead{Baseline} &
\colhead{API$^a$} &
\colhead{On-source} &
\colhead{Gain} \\
\colhead{} &
\colhead{Dates} &
\colhead{} &
\colhead{} &
\colhead{Lengths [km]} &
\colhead{rms [$\degr$]} &
\colhead{time [min]} & 
\colhead{calibrator(s)}
}
\startdata
HD 377 & 2014 Jul 3 & D & 27 & $0.04-1.31$ & 6.6 & 64.8 & J0011$+$0823  \\
 & 2014 Jul 10 & D & 26 &$0.04-1.31$ & 5.6 & 62.1 & \\
49 Ceti & 2014 Jul 13 & D & 27 & $0.04-1.31$ & 5.8 & 63.3 & J0132$-$1654  \\
 & 2014 Jul 14 & D & 25 & $0.04-1.31$ & 3.4 & 66.4 & \\
HD 15115 & 2014 Jul 8 & D & 26 & $0.04-1.31$ & 3.0 & 65.7 & J0224$+$0659  \\
 & 2014 Jul 11 & D & 26 & $0.04-1.31$ &3.1 & 65.6 & \\
HD 61005 & 2014 Sep 19 & DnC & 27 & $0.04-2.11$ & 1.5 & 62.0 & J0747$-$3310 \\
 & 2014 Sep 20 & DnC & 27 & $0.04-2.11$ & 3.3 & 62.2 & \\
HD 104860 & 2014 Aug 30 & D & 27 & $0.04-1.31$ & 5.3 & 61.4 & J1220$+$7105  \\
 & 2014 Aug 30 & D & 27 & $0.04-1.31$ & 4.6 & 61.2 & \\
HD 141569 & 2014 Jun 6 &  D & 25 & $0.04-1.31$ & 4.0 & 61.5 & J1557$-$0001  \\
AU Mic & 2013 May 9 & DnC & 27 & $0.04-2.11$ & 4.1 & 49.7 & J2101$-$2933 \\
 & 2013 May 11 & DnC & 27 & $0.04-2.11$ & 2.8 & 49.9 & \\
 & 2013 Jun 21 & C & 27 & $0.05-3.38$ & 1.3 & 50.0 & \\
\enddata
\tablecomments{$^\text{a}$ Measure of the tropospheric contribution to the interferometric phase determined by the Atmospheric Phase Interferometer (API), an interferometer comprised of two 1.5-m antennas separated by 300~m, observing an 11.7 GHz beacon from a geostationary satellite.
}
\label{tab:obs}
\end{deluxetable}

\begin{deluxetable}{l c c c c}
\tablecolumns{5}
\tabcolsep0.1in\footnotesize
\tabletypesize{\footnotesize}
\tablewidth{0pt}
\tablecaption{Results of the VLA Observations}
\tablehead{
\colhead{Source} & 
\colhead{Beam Size$^a$} & 
\colhead{Beam P.A.$^b$} & 
\colhead{$F_\text{9mm}$} & 
\colhead{rms Noise} \\
\colhead{} & 
\colhead{[$\arcsec$]} &
\colhead{[deg.]} &
\colhead{[$\mu$Jy]} &
\colhead{[$\mu$Jy/beam]}
}
\startdata
HD 377 & $2.8\times2.4$ & 21.2 & $<13.1$ & 4.4 \\
49 Ceti & $3.8\times2.6$ & 339.6  & 25.1 & 5.5  \\
HD 15115 & $3.0\times2.4$ & 314.0 & 12.8 & 4.1  \\
HD 61005 & $2.6\times2.0$ & 45.6 & 57.3 & 8.6  \\
HD 104860 & $3.1\times2.3$ & 75.6 & 14.0 & 3.5  \\
HD 141569 & $3.0\times2.4$ & 338.6 & 85.0 & 5.1 \\
AU Mic & $3.1\times2.8$ & 41.0 & $>60.8$ & 5.2  \\
\enddata
\tablecomments{$^\text{a}$ Beam size determined with natural weighting\\ 
$^\text{b}$ Beam position angle measured from east of north
}
\label{tab:results}
\end{deluxetable}

\begin{deluxetable}{l c c c c c c c c c c}
\tablecolumns{11}
\tabcolsep0.05in\scriptsize
\tabletypesize{\scriptsize}
\tablewidth{0pt}
\tablecaption{Grain Size Distribution Slopes ($q$)}
\tablehead{
\colhead{Source} & 
\colhead{(Sub)mm} & 
\colhead{(Sub)mm} &
\colhead{$F_\text{mm}$} & 
\colhead{Ref.$^a$} & 
\colhead{$\alpha_\text{mm}$} &
\colhead{$R_\text{dust}^\text{b}$} &
\colhead{Ref.$^a$} &
\colhead{$T_\text{dust}^\text{c}$} &
\colhead{$\alpha_\text{Pl}$} &
\colhead{$q$} \\
\colhead{} & 
\colhead{$\lambda$ [mm]} & 
\colhead{Instr.} & 
\colhead{[mJy]} &
\colhead{} &
\colhead{} &
\colhead{[AU]} &
\colhead{} &
\colhead{[K]} &
\colhead{} &
\colhead{}
}
\startdata
HD 377 & 0.87 & SMA & $3.5\pm1.0$ & 1 & $>2.39$ & 50 & 1 & $44$ & $1.92\pm0.02$ & $>3.26$ \\
49 Ceti & 0.85 & ALMA & $17\pm3$ & 2 & $2.76\pm0.11$ & 120 & 2 & $55$ & $1.94\pm 0.01$ & $3.46\pm0.08$ \\
HD 15115 & 1.3 & SMA & $2.6\pm0.6$ & 3 & $2.75\pm0.15$ & 110 & 15 & $45$ & $1.92\pm0.02$ & $3.46\pm0.10$ \\
HD 61005 & 1.3 & SMA & $7.2\pm0.3$ & 4 & $2.49\pm0.08$ & 70 & 4 & $30$ & $1.91\pm0.03$ & $3.32\pm0.06$ \\
HD 104860 & 1.3 & SMA & $4.4\pm1.1$ & 1 & $3.08\pm0.23$ & 110 & 1 & $28$ & $1.90^{+0.02}_{-0.04}$ & $3.64\pm0.15$ \\
HD 141569 & 0.87 & ALMA & $3.78\pm0.45$ & 5 & $1.63\pm0.06$ & 30 & 5 & $109$ & $1.97\pm0.02$ & $2.84\pm0.05$ \\
AU Mic & 1.3 & ALMA & $7.14\pm0.15$ & 6 & $<2.46$ & 20 & 6 & $25$ & $1.90^{+0.03}_{-0.07}$ & $<3.31$ \\
\\
\hline
\\
q$^1$ Eri & 0.87 & APEX & $39.4\pm4.1$ & 7 & $2.94\pm0.10$ & 85 & 7 & $33$ & $1.88^{+0.02}_{-0.04}$ & $3.59\pm0.08$ \\
$\epsilon$ Eri & 1.3 & SMA & $17.2\pm5.0$ & 8 & $>2.39$ & 64 & 8 & $27$ & $1.89^{+0.03}_{-0.06}$ & $>3.28$ \\
$\beta$ Pic & 0.87 & ALMA & $60\pm6$ & 9 & $2.81\pm0.10$ & 85 & 9 & $52$ & $1.93\pm0.01$ & $3.49\pm0.08$ \\
HD 95086 & 1.3 & ALMA & $3.1\pm0.18$ & 10 & $2.37\pm0.15$ & 90 & 16 & $50$ & $1.93\pm0.01$ & $3.24\pm0.10$ \\
HD 107146 & 1.25 & ALMA & $12.5\pm1.3$ & 11 & $2.55\pm0.11$ & 60 & 11 & $30$ & $1.90^{+0.05}_{-0.02}$ & $3.36\pm0.07$  \\
AK Sco & 1.3 & ALMA & $32.65\pm0.07$ & 12 & $2.62\pm0.03$ & 14 & 12 & $95$ & $1.97\pm0.01$ & $3.36\pm0.04$ \\
HD 181327 & 1.3 & ALMA & $7.5\pm0.1$ & 13 & $2.38\pm0.05$ & 90 & 17 & $60$ & $1.94\pm0.01$ & $3.24\pm0.05$  \\
Fomalhaut & 0.85 & SCUBA & $27.0\pm3.0$ & 14 & $2.70 \pm 0.17$ & 135 & 18 & $48$ & $1.91\pm0.02$ & $3.44\pm0.11$ \\
\enddata
\tablecomments{$^a$ References: 1) \cite{stee15}, 2) Hughes et al. (in prep), 3) \cite{mac15a}, 4) \cite{rica13}, 5) White et al. (in prep), 6) \cite{mac13}, 7) \cite{lis08}, 8) \cite{mac15b}, 9) \cite{den14}, 10) Su et al. (2016, in prep), 11) \cite{ric15}, 12) \cite{cze15}, 13) Marino et al. (in prep), 14) \cite{ric12}, 15) \cite{mac15a}, 16) \cite{su15}, 17) \cite{stark14}, 18) \cite{bol12}\\
$^b$ Uncertainty on $R_\text{dust}$ assumed to be 20\%\\
$^c$ Uncertainty on $T_\text{dust}$ assumed to be $10\%$
}
\label{tab:qvalues}
\end{deluxetable}

\end{document}